# Electronic stopping for protons and α-particles from first principles electron dynamics: The case of silicon carbide

Dillon C. Yost and Yosuke Kanai*
*Department of Chemistry, University of North Carolina at Chapel Hill, Chapel Hill, NC 27516, USA*
(Dated: May 1, 2018)

We present the first-principles determination of electronic stopping power for protons and α-particles in a semiconductor material of great technological interest: silicon carbide. The calculations are based on non-equilibrium simulations of the electronic response to swift ions using real-time time-dependent density functional theory (RT-TDDFT). We compare the results from this first-principles approach to those of the widely used linear response formalism and determine the ion velocity regime within which linear response treatments are appropriate. We also use the non-equilibrium electron densities in our simulations to quantitatively address the long-standing question of the velocity-dependent effective charge state of projectile ions in a material, due to its importance in linear response theory. We further examine the validity of the recently proposed centroid path approximation recently proposed for reducing the computational cost of acquiring stopping power curves from RT-TDDFT simulations.

## I. INTRODUCTION

Understanding the stopping process of highly energetic ions in condensed matter systems has great implications in modern technologies ranging from nuclear fission/fusion reactors [1], to semiconductor devices for space missions [2], to cancer therapy based on ion beam radiation [3]. The kinetic energy of irradiating energetic ions is dissipated in a material in the stopping stage, a fundamental process in which deposited energy becomes available for inducing structural transitions through various mechanisms. Conceptually, the stopping stage is divided into two regimes, depending on the type of excitation produced [4]: At low ion velocities, the dominant effect is nuclear stopping, which primarily results in lattice excitations and nuclei displacements. At higher velocities (typically > keV), the relevant excitations are electronic, hence the term 'electronic stopping'. The average rate of energy transfer from the ion to the target material is generally measured with respect to the unit distance of projectile ion movement, and this is referred to as stopping power.

Ever since the phenomenon of electronic stopping was discovered, a number of approximated analytical models have been developed: the classical Coulomb scattering formulas of Rutherford [5], Thomson [6], and Darwin [7], the quantum-mechanical perturbation approach by Bethe [8], electron gas models by Fermi and Teller [9], and the dielectric formalism treatment by Lindhard (see [10-12] and references therein). Non-perturbative calculations (necessary to model, e.g., the Barkas effect [13] and so-called $Z_1$ oscillations) of electronic stopping in the uniform electron gas started in the 1980s by Echenique and co-workers [14,15] with the advent of density functional calculations and their time-dependent counterparts [16]. For historical reviews of theoretical approaches to electronic stopping see Refs. [17,18]. Today, perhaps the most widely used approach is the linear response formalism, originating with Bethe, but also used by Lindhard and others. In the framework of linear response theory (consequently, the projectile ion is assumed to have a fixed charge Z with no velocity dependence), the stopping power can be expressed in a mathematically closed form [19]:

$$S(v) = \frac{4\pi Z^2}{v^2} L(v) \quad (1)$$

where $v$ is the projectile ion velocity, and $L(v)$ is a velocity-dependent quantity called the stopping logarithm. This quantity is given in terms of either mean excitation energy of the target material in Bethe theory [8], or as the energy/wave-vector dependent dielectric response function in Lindhard's formula [10,11]. Note that the mean excitation energy can be obtained from the optical limit of dielectric response function [20,21] or from electronic structure calculations [22,23], whereas Lindhard's approach requires a full dielectric function:

$$L(v) = \frac{1}{2\pi^2} \int_0^{qv} \omega \, d\omega \int_0^\infty \frac{dq}{q} Im\left(\frac{1}{\varepsilon(q,\omega)}\right) \quad (2)$$

where $\varepsilon$ is the macroscopic dielectric function of frequency $\omega$ and wavelength $q$.

In the last few decades, both rapidly advancing high-performance computers and modern electronic structure methods have made it possible to obtain key parameters in the analytical models directly from first principles theory [24-27]. Parameter-free methods can go significantly beyond analytical models because they provide detailed information at the atomistic level, allowing one to study the specific influences of defects, surfaces, or even the nature of electronic excitations involved in the stopping process. However, a fully atomistic first-principles calculation of electronic stopping for a wide range of projectile velocities, especially around the maximum of the electronic stopping curve, has remained elusive. The possibility of quantitatively describing the interaction of projectile atoms with the electronic *and* ionic system of the host material entirely within first-principles calculations has come within reach [28,29]. These advances for realistic materials rely on non-perturbative, real-time, time-dependent density functional theory (RT-TDDFT) [16]. In a recent study [30], we demonstrated an accurate calculation of the electronic stopping power curve for protons and α-particles in a representative metallic system of aluminum, for which practical approximations within RT-TDDFT, such as the exchange-correlation potential, are thought to be satisfactory.

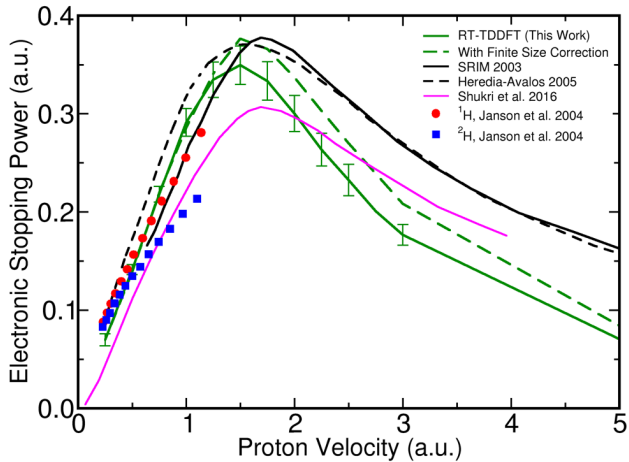 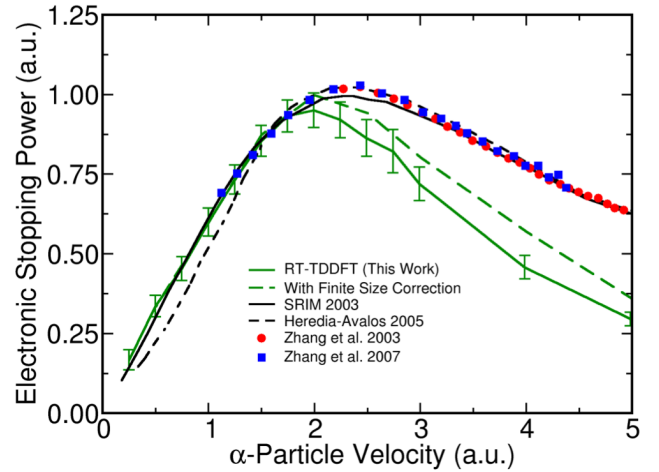

FIG. 1. Electronic stopping power as a function of the velocity of a proton projectile in 3C-SiC. The solid black curve represents the empirical SRIM model [52]. The dashed black curve represents a dielectric formalism calculation with an empirically fitted dielectric function [48]. Red circles and blue squares show experimental data for protons and deuterium ions, respectively [47]. The pink curve represents results from the dielectric response formalism using LR-TDDFT for calculating the dielectric matrix [26]. The solid green curve corresponds to the values we obtained by calculating the non-equilibrium response using RT-TDDFT. The error bars represent standard deviations for the path distribution. The dashed green curve shows the RT-TDDFT results with an added finite size error correction.

FIG. 2. Electronic stopping power as a function of the velocity of an α-particle projectile in 3C-SiC. The solid black curve represents the empirical SRIM model [52]. The dashed black curve represents a dielectric formalism [48] calculation with an empirically fitted dielectric function. Red circles and blue squares show experimental data for α-particles over two velocity ranges [53,54]. The solid green curve corresponds to the values we obtained by calculating the non-equilibrium response using RT-TDDFT. The error bars represent standard deviations for the path distribution. The dashed green curve shows the RT-TDDFT results with an added finite size error correction.

In light of this encouraging result, we apply the methodology to the representative semiconductor material silicon carbide (SiC). SiC has garnered attention for fusion and advanced fission energy applications due to its ability to retain important physical and chemical properties when exposed to extreme particle radiation [31-34]. Also, SiC has potential for use in semiconductor nuclear radiation detectors due to its ability to withstand radiation-induced damage better than conventional semiconductor materials such as silicon and germanium [35]. These applications make SiC a scientifically and technologically relevant case for studying electronic stopping in semiconductors. In addition to the non-empirical determination of the electronic stopping power for protons and α-particles from first-principles simulations, we address the long-standing question of the effective charge state of the projectile ion and the related issue of the extent to which a linear-response formalism can be applied over different velocity regimes. Also, we examine the validity of the recently used centroid path approximation for calculating stopping power [36,37].

## II. THEORETICAL METHOD

The simulation methods employed in this work closely follow that described by Schleife et. al. [38,39] involving a real-time propagation approach within time-dependent density functional theory. For all simulations, the PBE exchange-correlation functional was used [40] within the adiabatic approximation [41,42]. In the current method, a plane-wave pseudopotential scheme is used in solving the time-dependent Kohn-Sham equations in which a real, swift ion (proton or α-particle) is responsible for the time-dependence of the external potential acting on the electronic system. Hamann-Schluter-Chiang-Vanderbilt norm-conserving pseudo-potentials were used for all atoms [43], including the projectile ions. We employ our recently-developed, highly parallelized implementation of RT-TDDFT [38,39] in the Qb@ll branch of the Qbox code [44,45].

In this work, we use a simulation cell consisting of 216 atoms (864 electrons) in a cubic super-cell (lattice constant 4.36 Å) of 3C-SiC, the zinc-blende polytype of silicon carbide. Despite the large simulation cell we employ, the electronic stopping power curve is not completely converged with respect to the simulation cell size. Unfortunately, it is not possible to achieve the strictest convergence at the present time because of the large computational cost associated with these RT-TDDFT simulations in obtaining the ensemble average. Instead, we used a representative projectile ion path (i.e. centroid path as discussed later) to estimate the finite size correction using a much larger simulation cell as discussed in detail in Supplemental Material. For the remainder of the work, electronic stopping power curves are shown with and without the finite size error correction. The gamma point was used in sampling of the Brillouin zone, and it was found to be sufficient for convergence by comparing with calculations with 4 k-points as shown in Supporting Information. A plane-wave energy cutoff of 50 Rydberg was used. In this work, we did not use a single, long, re-entering projectile path for obtaining an ensemble average as done previously for metallic systems [30]. Instead, we directly obtained the ensemble average using 10 independent projectile ion paths that are determined via a random number generator, and there are no constraints on the impact parameters. Thus, there are rare instances in which the projectile ion penetrates into the pseudo-potential spheres of other atoms. The positions of all atoms, except the projectile ion, are held fixed in the simulation cell while the electronic system evolves in response to the time-dependent potential due to the projectile ion. The non-equilibrium simulations yield the electronic energy increase as a function of the projectile displacement for a specific ion velocity. We then apply a baseline fitting with asymmetric least squares fitting proposed by Eilers et al. [46] in order to acquire a linear regression. This slope represents the energy derivative that can be used to calculate the electronic stopping power via the following equation:

$$S(v) = \langle \frac{dE[\rho(r;t)]}{dx} \rangle_v, \quad (3)$$

where $E$ is the time-dependent electronic energy and $\rho(r;t)$ is the non-equilibrium electron density [30], and $x$ is the projectile ion position.

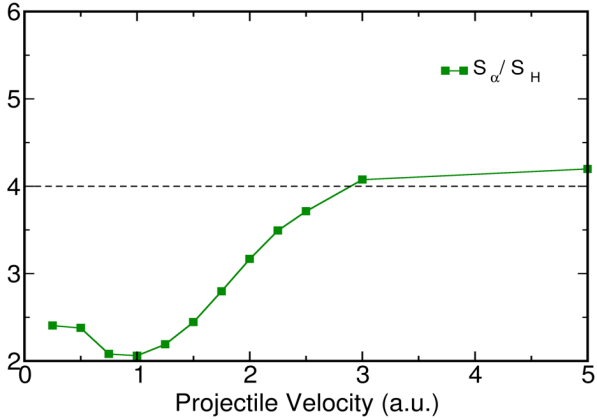

FIG. 3. Ratio of the calculated electronic stopping power for α-particle to the electronic stopping power for proton, as a function of the projectile ion velocity. The dashed line indicates $(Z_\alpha/Z_H)^2 = 4$ one would obtain using linear response theory by assuming fully-ionized projectile ions.

## III. RESULTS AND DISCUSSION

### A. Electronic stopping power for proton and α-particle

Janson and co-workers [47] conducted ion-implantation experiments to acquire low-velocity electronic stopping power for $^1$H and $^2$H ions using time-of-flight (TOF) techniques. The experimental data shown in Figure 1. was obtained by removing the nuclear stopping power component from their experimental measurements. The analytical model by Heredia-Avalos and co-workers [48] employs the dielectric response formulation using a Mermin-type dielectric function [49] together with a modified Brandt-Kitagawa model [50,51] for the effective charge state of the proton. The dielectric function was obtained by fitting to the experimental spectrum of the energy loss function in the optical limit ($q = 0$). Additionally, SRIM 2003 provides empirically fitted data from a combination of experimental results for electronic stopping in Si and C [52]. As can be seen in Figure 1, the analytical model by Heredia-Avalos, et al. [48] and the SRIM model are in rather good agreement with each other, especially for higher velocities beyond the stopping power maximum.

As an alternative to using a model dielectric function, the dielectric formalism can be cast in terms of the microscopic dielectric function, which can be computed using modern first-principles electronic structure calculations [24,25]. Recently, Shukri and co-workers [26] employed such a dielectric response formalism:

$$S(v) = \frac{4\pi Z^2}{N_k \Omega v} \sum_q^{BZ} \sum_G Im\{\varepsilon_{G,G}^{-1}(q,\omega)\} \frac{v\cdot(q+G)}{|q+G|^2} \quad (4)$$

where $N_k$ is the number of $k$ points used in the Brillouin zone, $\Omega$ is the volume of the unit cell, $q$ is a lattice vector in the first Brillouin zone, $G$ are reciprocal lattice vectors, $v$ is the ion velocity and the microscopic dielectric matrix was calculated using linear response TDDFT with the adiabatic local density approximation (ALDA). The energy dependence is given by $\omega = v \cdot (q + G)$, and their result for the cubic 3C-SiC is also shown for comparison in Figure 1.

In Figure 1, our RT-TDDFT simulation results for protons in SiC are shown with and without the finite-size error correction as discussed in Theoretical Method section. Note that the correction becomes appreciable for the ion velocities beyond the stopping power peak. Our RT-TDDFT simulation results show good agreement with the available experimental data and the empirical models for the low-velocity regime (v < 1.5 a.u.). Additionally, the position and magnitude of the stopping power peak is in good agreement with the empirical models, showing the stopping power maximum at v = ~1.5 a.u., which is close to the result by Heredia-Avalos, et al. [48]. However, for the higher velocities (v > 2 a.u.), the RT-TDDFT simulation results yield stopping powers that are significantly lower (~50%) than those given by the empirical models even when the finite size error is taken into account. The linear response formulation is expected to become more accurate with the increasing velocity, and the noticeable difference between our RT-TDDFT result and the linear response result using the TDDFT microscopic dielectric matrix [26] is notable even for the high velocity regime. Part of the disagreement between these two first-principles approaches stem from the use of the PBE XC approximation in our RT-TDDFT simulations and the use of the LDA XC approximation in calculating TDDFT microscopic dielectric matrix [26]. Indeed, when we calculate the electronic stopping power using LDA in RT-TDDFT simulations, the resulting stopping power is larger by as much as 18 % for the velocities above ~1.5 a.u. (see Supplemental Material). Another source of the disagreement might come from the neglect of core electron excitations in our RT-TDDFT simulations. From an earlier work on silicon [26], neglecting excitations of 2s and 2p electrons is likely to result in a slight underestimation even in SiC for velocities beyond the stopping power maximum, but not enough to fully explain the disagreement. Given that our RT-TDDFT simulation uses the approximated finite-size error correction and the linear response result by Shukri et al. uses an extrapolation scheme [26] (because achieving strict convergence is not possible at the present time), the observed disagreement, even for high velocity regimes, calls for a systematic examination of both approaches in a future work. At the same time, we note that for a simpler metallic case of aluminum, these two first-principles approaches have been shown to agree quite well for the case of protons, as previously discussed [26,30].

Figure 2. shows our RT-TDDFT simulation result in comparison with experimental measurements for α-particles. For α-particles, Zhang et al. used a TOF setup to determine electronic stopping power of He ions in SiC over a wide velocity range [53,54]. There is excellent agreement between our result and the experiments for the velocity range below the stopping power maximum. However, for higher velocities, our results are significantly lower than the experimental stopping power data even when the finite-size error is taken into account. The disagreement with the experimental measurements is indicative of underlying approximations in RT-TDDFT simulations, specifically the XC and adiabatic approximations.

Comparing the stopping power curves for these two different projectile ions, proton and α-particle, (Fig 1. and Fig 2.), there is a shift in the stopping power curve maximum going from proton to α-particle: The peak of the proton stopping power curve is located at v = ~1.5 a.u., whereas the peak for the α-particle stopping power curve is at v = ~2.0 a.u. Such a peak shift cannot be obtained by employing a linear response model. Within linear response theory, the stopping power has a quadratic dependence on the projectile ion charge since the stopping logarithm (Eq. 2) depends only on the target medium. Considering the proton and α-particle curves, the ratio $S_\alpha(v)/S_H(v) = 4$ represents the validity of the linear response theory at a specific velocity, assuming fully-ionized projectiles.

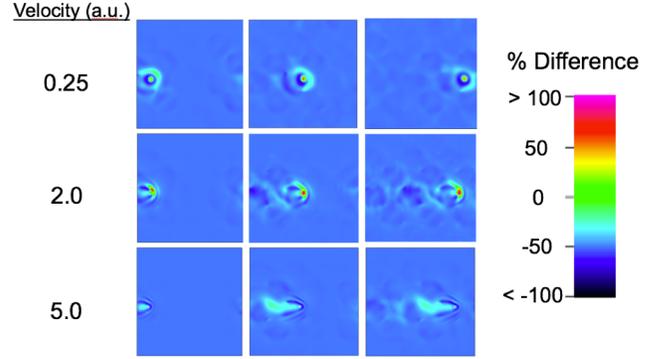

FIG. 4. Response-normalized density differences, $\Delta\rho_{\alpha-H}$, between α-particle and proton partway through the simulations. Blue and red indicate negative and positive deviations, respectively, from the induced density predicted by linear response theory. Volume slices parallel to the projectile ion path are shown. For clarity, atoms are not shown.

The ratio, $S_\alpha(v)/S_H(v)$, from our RT-TDDFT results is plotted in Figure 3. as a function of the ion velocity. The ratio of the stopping power curves for α-particles to that for protons approaches 4 for velocities larger than v = ~3 a.u., which is well beyond the stopping power maxima for both protons and α-particles. This corroborates the notion that additional higher-order Z corrections and/or effective ion charge models are necessary for the linear response theory to correctly capture the stopping power maximum, despite the inconvenience associated with having more empirical parameters [55,56].

The observed difference in stopping power curves between protons and α-particles is directly related to the difference in the non-adiabatic forces on the projectile ion [57], and it is informative to analyze the spatial dependence of non-equilibrium electron density beyond what is described by linear response theory. In the electronic stopping of ions, induced electron density is proportional to the electronic stopping power. To this end, the *response-normalized* density difference is calculated as follows:

$$\Delta\rho_{\alpha-H}(t) = \frac{\rho_\alpha(t)/2 - \rho_H(t)}{\rho_H(t)} \times 100\% \quad (5)$$

where $\rho_\alpha(t)$ and $\rho_H(t)$ are the time-dependent electron densities for the α-particle and proton cases, respectively. The factor of 2 takes into account the fact that the induced electron density response is twice as large with α-particle within linear response theory. In plotting the response-normalized density difference, as in Figure 4, one can directly observe the deviations of linear response theory from the RT-TDDFT simulation results. Figure 4 shows $\Delta\rho_{\alpha-H}(t)$ at a representative instant of time for three different velocities, representing the low-velocity region (v = 0.25 a.u.), the velocity region near peak stopping power (v = 2.0 a.u.), and the high-velocity region (v = 5.0 a.u.). Green color in Figure 4 would represent the response of electron density that is consistent with linear response treatment. As can been seen, the density response for the α-particle case is much lesser in comparison for most regions except the immediate vicinity of the projectile ion.

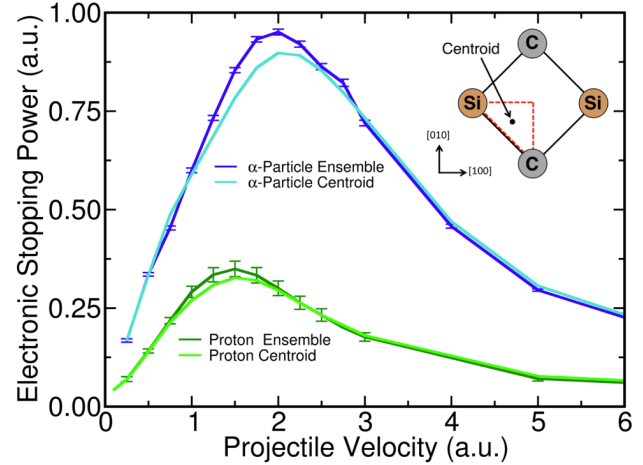

FIG. 5. Electronic stopping power calculated from the ensemble average of RT-TDDFT simulations for α-particle (darker blue) and proton (darker green) in SiC. Calculation results using the centroid path approximation are shown for α-particle (light blue) and proton (light green). Shown in the upper right corner is a schematic indicating the centroid path in 3C-SiC relative to atomic positions of C and Si.

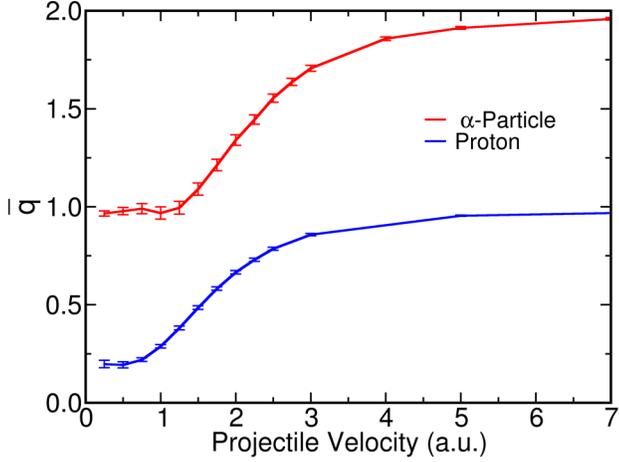
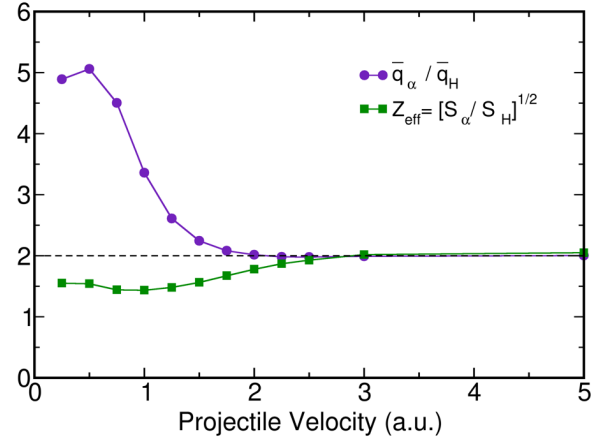

FIG. 6. Velocity-dependent mean steady-state charges, $\bar{q}$, for α-particle (red) and proton (blue) in SiC. Error bars indicate standard deviations of the mean. The dependence of the charge on impact parameters is shown in Supplemental Material.

FIG. 7. Ratio of mean steady-state charge (purple) between α-particle and proton over a range of projectile velocities. The green curve corresponds to $Z_{eff}$, the effective scaling factor calculated from the stopping power ratios between α-particle and proton. Within linear response theory, the ratio $Z_\alpha/Z_H$ yields 2 as indicated by dashed line if fully-ionized projectiles are assumed.

### B. Examining the centroid path approximation

The instantaneous energy loss rate of the projectile ion in a condensed matter system often depends strongly on the specific path taken by the ion and its proximity to atoms and bonds over the course of the trajectory. In order to obtain the electronic stopping power in real materials like SiC, an ensemble average over numerous projectile paths needs to be taken until satisfactory convergence is reached. Unfortunately, numerous expensive RT-TDDFT simulations are necessary to obtain an accurate ensemble average, making it a computationally demanding procedure. In an attempt to reduce this computational expense and still acquire accurate results, Ojanperä et al. [36] posited that for the case of a symmetric two-dimensional system like graphene, a geometric centroid path could be used to approximate the ensemble average of projectile paths through the graphene surface. However, a thorough analysis of this approximation for different ion velocities and different materials has not been reported. In order to examine how well this approach works for 3D crystalline materials, like SiC, over a wide velocity range, we compared the calculated electronic stopping power curves from the 3C-SiC centroid path to the ensemble average from 10 random paths.

The centroid path was determined by considering a 2D orthographic perspective of one channel of 3C-SiC. This perspective is deconstructed into its irreducible representation consisting of a triangle with a silicon atom, a carbon atom, and the channeling point as vertices (see Figure 5). The centroid of this triangle is given by the intersection of the triangle's three medians. Finally, the direction of the centroid path is given the path along the [001] crystal direction that passes through the centroid point. This procedure can easily be abstracted to other materials with symmetric crystal structures. Figure 5. shows the centroid path from a 2D orthographic perspective of a 3C-SiC channel, where the ion path is through the centroid along the [001] direction.

Figure 5. shows the stopping power curves from both the centroid path and the ensemble average of 10 random projectile paths. Except for some differences near the stopping power peak, the centroid path stopping power is in remarkably good agreement with the ensemble average. This implies that the electronic response along the centroid path is close to the average of the electronic responses in 3C-SiC. In recent RT-TDDFT simulation work by Ullah *et al.* [58] a systematic investigation of impact parameters along channeling paths in bulk cubic Germanium showed that the electronic stopping power can be related to the average density along the trajectory. Our results also support this finding, and they support the notion that the average electron density along the centroid path result in a stopping power that is in good agreement with the ensemble average of random paths. Thus, these results appear to support the proposition put forth by the empirical Bragg's additivity rule that the electronic stopping power is mainly proportional to the average density of electrons [59]. However, the results also imply that this empirical rule becomes less acceptable near the peak of the stopping power curve (6.5% and 5.5% underestimations at the stopping power maximum for proton and α-particle, respectively). Shukri et al. [26] found that Bragg's additivity rule is only effectual at higher velocities (v > 2 a.u.) for SiC, and that the deviation between Bragg's rule stopping power and their calculated stopping power was as large as 15% for low to moderate ion velocities. It appears that the centroid path approximation is able to partly capture details of the chemical bonds in the target material (so-called "bond effects" [22]), which are completely missed when employing Bragg's additivity rule.

### C. Effective charge state and Linear Response Theory

Within linear response theory, one starts by considering a particle with a fixed charge interacting with a material. A natural question is whether or not this fixed charge is different from the ion charge in vacuum, and also the extent to which the charge state depends on the inhomogeneous electron density in real materials. As early as the 1920s, experiments by Rutherford [60] showed the presence of singly-charged helium atoms in the beam of α-particles emerging from a penetrated material. Not only were singly charged $He^+$ observed, but also it was found that the ratio of $He^+$ to $He^{2+}$ ions increases at lower ion velocities. Sequences of electron capture and loss events yield the mean steady-state charge, $\bar{q}$, on the projectile ion [61,62]. The mean steady-state charge has a velocity-dependence, and it varies widely between solid and gaseous stopping media [63], with solids giving rise to more full ionization. In addition to this mass-density effect, the velocity dependence of the charge state has also been widely studied, and various theoretical descriptions exist in the literature. The commonly used Brandt-Kitagawa theory [50,51], for example, models the charge state as a function of the scaled velocity based on the Thomas-Fermi model. Clearly, an important aspect in applying the linear response theory formalism for calculating stopping power is the question of whether the use of an effective charge for the swift ion can better represents the electronic stopping power curve.

As noted by other authors [64], terminological confusion with regards to the use of the "effective charge" of a projectile has pervaded literature due to the term's two different definitions: The original concept of "effective charge" proposed by Bohr and others [61,65] referred to the real steady-state charge of the ion (i.e., the $\bar{q}$, mean steady-state charge value mentioned above). It was not until later [66] that the same terminology was used to describe a related, but distinctively different concept: effective charge state for the projectile ion was then defined such that $Z_{eff} = [S_{ion}/S_H]^{1/2}$ is satisfied. The relation between these two quantities remains unclear, and they cannot be assumed to be equivalent. For low velocities, projectile ions are usually assumed to be nearly neutral, giving rise to the deduction that $Z_{eff} > \bar{q}$ for low velocities. However, for high velocities, the relation has been widely debated, with some experiments on solid targets indicating $Z_{eff} > \bar{q}$, and experiments on gaseous targets indicating $Z_{eff} \cong \bar{q}$ [62].

In principle, all necessary information for calculating the mean steady-state charge, $\bar{q}$, on the projectile ion is contained in our RT-TDDFT simulations. However, a sensible partitioning scheme for non-equilibrium electron density is needed to quantify the electron charge belonging to the projectile ion. We presently employ the Voronoi analysis [67] using analysis code by Henkelman et al. [67]. Forty equally spaced electron density "snapshots" at different times were taken for each projectile velocity traveling along the centroid path in the RT-TDDFT simulations. Next, induced electron densities were calculated by subtracting the 3C-SiC ground state electron density from the non-equilibrium electron densities at the different times. These induced electron densities give a spatial representation of where electron density is accumulating in the simulation cell, and where it is being depleted. Finally, the Voronoi analysis is performed on the induced electron densities to quantify the charge within the projectile ion's Voronoi cell at different positions in the trajectory. While partitioning schemes based on electron density topology (e.g. Bader decomposition) are commonly used for the ground-state electronic density, the present problem lends itself better to the above approach using geometry-based Voronoi analysis [67] because its partitioning scheme satisfies two key criteria that are required for our analysis: First, the Voronoi partitioning criterion is not affected by the projectile ion velocity, which is necessary in order to have a consistent definition for all ion velocities. Second, the partitioning scheme approaches the correct limit (fully ionized projectile ion) in the high velocity limit. The Voronoi cell of a given atom is defined as the region of space closer to the given atom than to any other atom. In crystalline materials, the Voronoi cell is equivalent to the Wigner-Seitz cell. It is this geometric criterion that ensures that the size and shape of the ion's Voronoi cell depends only on the position of the projectile ion, not its velocity. Another advantage of Voronoi analysis is that it allows us to quantify the charge of the projectile ion throughout the entire trajectory. This can give insight into the dynamics of the charge capture and loss process, and it allows for the calculation of a mean steady-state charge.

Figure 6. shows the mean steady-state charge of the projectile ion as a function of the ion velocity. As its velocity increases, the projectile approaches a fully ionized state ($\bar{q} = Z$). The Voronoi partitioning scheme yields this known exact

behavior in the limit of high velocities. For the lowest velocity simulated, v = 0.25 a.u., the ions do not approach complete neutrality. Instead, the proton and α-particle approach ion charges of approximately +0.25 and +1.00, respectively. Interestingly, a recent RT-TDDFT simulation work by Zhao et al. [37] on two-dimensional boron-nitride and graphene sheets also shows that the α-particle acquires only about one electron or fewer for low velocities.

Returning to the question of the relationship between the concepts of the effective charge, $Z_{eff}$, and Bohr's original definition of effective charge, $\bar{q}$, it is interesting to examine the extent to which these two quantities become equivalent. Can scaling with the mean charge states rather than assuming fully ionized ion at all velocities remedy the shortcomings of linear response theory at low ion velocities? We examined $\bar{q}_\alpha/\bar{q}_H$ from our simulations in comparison to the calculated $[S_\alpha/S_H]^{1/2}$. While the two quantities are in excellent agreement for higher ion velocities v ≥ 3 a.u., there is a significant qualitative difference for v < 3 a.u. Even if other electron density partitioning schemes were considered, it is highly unlikely that $\bar{q}_\alpha/\bar{q}_H$ would yield the behavior observed for $[S_\alpha/S_H]^{1/2}$. Thus, for the present case of SiC, a non-empirical determination of $Z_{eff}$ in the context of linear response theory for predicting the stopping power curves for low ion velocities would not be feasible unless additional higher order perturbations were taken into account.

## IV. CONCLUSIONS

In this work, we presented first-principles calculations of electronic stopping power in cubic silicon carbide (SiC) for protons and α-particles, from non-equilibrium electron dynamics simulations based on real-time time-dependent density functional theory (RT-TDDFT). We have shown that the centroid path approximation [36,37] accurately reproduces electronic stopping power values of the ensemble average, while the agreement is worse for the velocities near the stopping power maximum. We have also quantified the velocity-dependent mean steady-state charges for protons and α-particles in SiC to examine the extent to which a linear response treatment can be applied. Our results indicate that linear response theory should be applicable for velocities larger than ~ 3 a.u. if the mean steady-state charges are used instead of assuming fully-ionized ions in SiC. While we have made significant progress toward an accurate determination of electronic stopping power and associated physical quantities, there remains much room for further investigation into the accuracy of first principles approaches. Our RT-TDDFT result and the linear response result by Shukri et al. [26] show a disagreement even for high velocities for the present case of the semiconductor SiC, and it calls for further examination of both approaches in a future work. This is at odds with the simpler case of aluminum for which these two first-principles approaches agree [26,30].

An important approximation in the present study is the exchange-correlation (XC) potential [68] used in our RT-TDDFT simulations. First, we adapted the adiabatic approximation such that the XC potential depends only on the instantaneous electron density, neglecting any potential memory effects. By using time-dependent current DFT to calculate the linear part of the stopping power in the low ion velocity limit for a homogeneous electron gas (i.e. friction coefficient), Nazarov and co-workers [69] have shown that the adiabatic approximation results in a negligibly small error for ions of low-Z elements like protons and α-particles. Second, the semi-local approximation, like GGA-PBE, used here for the XC potential could introduce non-negligible errors, especially since the dynamical charge transfer between the ion and target might be important. A future work will focus on exploring the dependence on XC approximation. The XC approximation can also be an important avenue of investigation for the threshold velocity at which the electronic stopping power diminishes [58]. Another more technical source of error is neglecting core electron excitations in our simulations. From an earlier work on silicon [26], error stemming from neglecting the excitations of 2s and 2p electrons would contribute to the non-negligible underestimation of the electronic stopping power for v ≥ 2 a.u. The role of core electrons in the high velocity regime will be studied in a future work.

## ACKNOWLEDGEMENTS

We acknowledge computational resources from the Argonne Leadership Computing Facility, which is a DOE Office of Science User Facility supported under Contract DE-AC02-06CH11357. This work is supported by the National Science Foundation under Grant No. CHE-1565714.

*ykanai@ad.unc.edu

[1] G. R. Odette and B. D. Wirth, in *Handbook of materials Modeling* (Springer, 2005), pp. 999.
[2] H. Zeller, Microelectronics reliability **37**, 1711 (1997).
[3] O. Obolensky, E. Surdutovich, I. Pshenichnov, I. Mishustin, A. Solov'Yov, and W. Greiner, Nuclear Instruments and Methods in Physics Research Section B: Beam Interactions with Materials and Atoms **266**, 1623 (2008).
[4] P. Sigmund, Bull. Russ. Acad. Sci. Phys. **72**, 569 (2008).
[5] E. Rutherford, The London, Edinburgh, and Dublin Philosophical Magazine and Journal of Science **21**, 669 (1911).

[6] J. J. Thomson, The London, Edinburgh, and Dublin Philosophical Magazine and Journal of Science **23**, 449 (1912).
[7] C. G. Darwin, The London, Edinburgh, and Dublin Philosophical Magazine and Journal of Science **23**, 901 (1912).
[8] H. Bethe, Annalen der Physik **397**, 325 (1930).
[9] E. Fermi and E. Teller, Physical Review **72**, 399 (1947).
[10] J. Lindhard, M. Scharff, and H. E. Schiøtt, *Range concepts and heavy ion ranges* (Munksgaard, 1963).
[11] J. Lindhard and A. Winther, *Stopping power of electron gas and equipartition rule* (Munksgaard, 1964).
[12] I. Nagy, I. Aldazabal, and M. Glasser, Journal of Physics B: Atomic, Molecular and Optical Physics **45**, 095701 (2012).
[13] W. H. Barkas, Annual Review of Nuclear Science **15**, 67 (1965).
[14] P. Echenique, R. Nieminen, J. Ashley, and R. Ritchie, Physical Review A **33**, 897 (1986).
[15] A. Arnau, M. Pealba, P. Echenique, F. Flores, and R. Ritchie, Physical review letters **65**, 1024 (1990).
[16] E. Runge and E. K. Gross, Physical Review Letters **52**, 997 (1984).
[17] C. Race, D. Mason, M. Finnis, W. Foulkes, A. Horsfield, and A. Sutton, Reports on Progress in Physics **73**, 116501 (2010).
[18] P. Echenique, F. Flores, and R. Ritchie, Solid State Physics **43**, 229 (1990).
[19] R. Cabrera-Trujillo and J. R. Sabin, Advances in Quantum Chemistry **45**, 1 (2004).
[20] B. J. McParland, in *Medical Radiation Dosimetry* (Springer, 2014), pp. 405.
[21] J. R. Sabin and J. Oddershede, Nuclear Instruments and Methods in Physics Research Section B: Beam Interactions with Materials and Atoms **44**, 253 (1990).
[22] J. R. Sabin and J. Oddershede, Nuclear Instruments and Methods in Physics Research Section B: Beam Interactions with Materials and Atoms **27**, 280 (1987).
[23] S. P. Sauer, J. R. Sabin, and J. Oddershede, Nuclear Instruments and Methods in Physics Research Section B: Beam Interactions with Materials and Atoms **100**, 458 (1995).
[24] I. Campillo, J. Pitarke, A. Eguiluz, and A. García, Nuclear Instruments and Methods in Physics Research Section B: Beam Interactions with Materials and Atoms **135**, 103 (1998).
[25] R. J. Mathar, J. R. Sabin, and S. Trickey, Nuclear Instruments and Methods in Physics Research Section B: Beam Interactions with Materials and Atoms **155**, 249 (1999).
[26] A. A. Shukri, F. Bruneval, and L. Reining, Physical Review B **93**, 035128 (2016).
[27] R. J. Mathar, S. B. Trickey, and J. R. Sabin, Advances in Quantum Chemistry **45**, 277 (2004).
[28] J. Pruneda, D. Sánchez-Portal, A. Arnau, J. Juaristi, and E. Artacho, Physical review letters **99**, 235501 (2007).
[29] R. Hatcher, M. Beck, A. Tackett, and S. T. Pantelides, Physical review letters **100**, 103201 (2008).
[30] A. Schleife, Y. Kanai, and A. A. Correa, Physical Review B **91**, 014306 (2015).
[31] Y. Katoh, N. Hashimoto, S. Kondo, L. L. Snead, and A. Kohyama, Journal of Nuclear Materials **351**, 228 (2006).
[32] Y. Katoh, L. L. Snead, C. H. Henager, A. Hasegawa, A. Kohyama, B. Riccardi, and H. Hegeman, Journal of Nuclear Materials **367-370**, 659 (2007).
[33] Y. Katoh, L. L. Snead, I. Szlufarska, and W. J. Weber, Current Opinion in Solid State and Materials Science **16**, 143 (2012).
[34] Y. Zhang, M. Ishimaru, T. Varga, T. Oda, C. Hardiman, H. Xue, Y. Katoh, S. Shannon, and W. J. Weber, Physical Chemistry Chemical Physics **14**, 13429 (2012).
[35] F. Ruddy, A. Dulloo, J. Seidel, S. Seshadri, and L. Rowland, Nuclear Science, IEEE Transactions on **45**, 536 (1998).
[36] A. Ojanperä, A. V. Krasheninnikov, and M. Puska, Physical Review B **89**, 035120 (2014).
[37] S. Zhao, W. Kang, J. Xue, X. Zhang, and P. Zhang, Journal of Physics: Condensed Matter **27**, 025401 (2014).
[38] A. Schleife, E. W. Draeger, V. M. Anisimov, A. A. Correa, and Y. Kanai, Computing in Science & Engineering **16**, 54 (2014).
[39] A. Schleife, E. W. Draeger, Y. Kanai, and A. A. Correa, The Journal of chemical physics **137**, 22A546 (2012).
[40] J. P. Perdew, K. Burke, and M. Ernzerhof, Physical review letters **77**, 3865 (1996).
[41] J. P. Perdew, K. Burke, and M. Ernzerhof, Phys Rev Lett **77**, 3865 (1996).
[42] N. T. Maitra, K. Burke, and C. Woodward, Phys Rev Lett **89** (2002).
[43] D. Vanderbilt, Phys Rev B **32**, 8412 (1985).
[44] F. Gygi, Qbox: a large-scale parallel implementation of First-Principles Molecular Dynamics, 2005.
[45] E. W. Draeger and F. Gygi, Lawrence Livermore National Laboratory Tech. Report. (2014).
[46] P. H. Eilers and H. F. Boelens, Leiden University Medical Centre Report (2005).
[47] M. S. Janson, M. K. Linnarsson, A. Hallén, and B. G. Svensson, Journal of applied physics **96**, 164 (2004).
[48] S. Heredia-Avalos, J. C. Moreno-Marín, I. Abril, and R. Garcia-Molina, Nuclear Instruments and Methods in Physics Research Section B: Beam Interactions with Materials and Atoms **230**, 118 (2005).
[49] N. D. Mermin, Physical Review B **1**, 2362 (1970).
[50] W. Brandt and M. Kitagawa, Physical Review B **25**, 5631 (1982).
[51] W. Brandt, Nuclear Instruments and Methods in Physics Research **194**, 13 (1982).

[52] J. F. Ziegler, Nuclear instruments and methods in physics research section B: Beam interactions with materials and atoms **219**, 1027 (2004).
[53] Y. Zhang and W. J. Weber, Applied physics letters **83**, 1665 (2003).
[54] Y. Zhang, J. Jensen, G. Possnert, D. A. Grove, I.-T. Bae, and W. J. Weber, Nuclear Instruments and Methods in Physics Research Section B: Beam Interactions with Materials and Atoms **261**, 1180 (2007).
[55] L. Porter, International journal of quantum chemistry **95**, 504 (2003).
[56] R. Cabrera-Trujillo, J. R. Sabin, and J. Oddershede, Physical Review A **68**, 042902 (2003).
[57] A. A. Correa, J. Kohanoff, E. Artacho, D. Sánchez-Portal, and A. Caro, Physical review letters **108**, 213201 (2012).
[58] R. Ullah, F. Corsetti, D. Sánchez-Portal, and E. Artacho, Physical Review B **91**, 125203 (2015).
[59] W. H. Bragg and R. Kleeman, The London, Edinburgh, and Dublin Philosophical Magazine and Journal of Science **10**, 318 (1905).
[60] E. Rutherford, Philosophical Magazine Series 6 **47**, 277 (1924).
[61] N. Bohr, *The penetration of atomic particles through matter* (I kommission hos E. Munksgaard, 1948), Vol. 18.
[62] H.-D. Betz, Reviews of Modern Physics **44**, 465 (1972).
[63] N. O. Lassen, Kgl. Danske Videnskab. Selskab, Mat.-fys. Medd. **26** (1951).
[64] N. R. Arista and A. n. F. Lifschitz, Advances in Quantum Chemistry **45**, 47 (2004).
[65] J. Neufeld, Physical Review **96**, 1470 (1954).
[66] L. C. Northcliffe, Studies in Penetration of Charged Particles in Matter **1133**, 353 (1964).
[67] G. Henkelman, A. Arnaldsson, and H. Jónsson, Computational Materials Science **36**, 354 (2006).
[68] C. A. Ullrich, *Time-Dependent Density-Functional Theory, Concepts and Applications* (Oxford Graduate Texts, 2011).
[69] V. U. Nazarov, J. M. Pitarke, Y. Takada, G. Vignale, and Y. C. Chang, Physical Review B **76**, 205103 (2007).

Supplemental Information for

"Electronic Stopping for Protons and α-particles from First-Principles Electron Dynamics: The Case of Silicon Carbide "

by Dillon C. Yost and Yosuke Kanai

## 1. k-point sampling of Brillouin Zone.

Convergence of k-point sampling in Brillouin zone was checked using a proton traveling along the centroid path with a velocity of 2 a.u. in the 216-atom supercell. Using 4 Monkhorst-Pack k-points instead of a single Gamma-point sampling, the stopping power would be higher only by 2.3%.

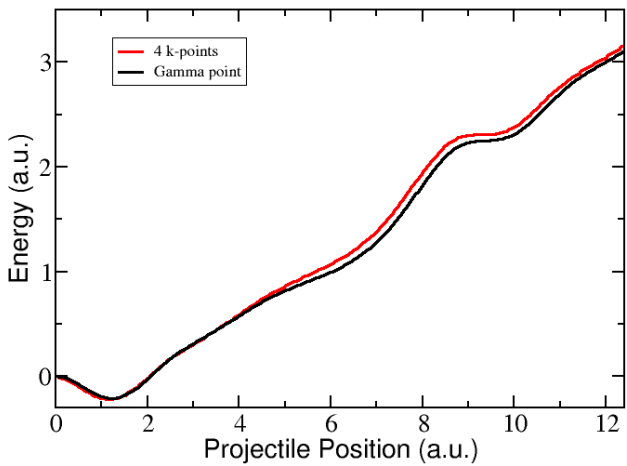

FIG. S1. Total electronic energy (relative to the ground state energy) as a function of the proton projectile displacement from the initial position. Simulation results for gamma point only (black) and 4 Monkhorst-Pack k-points (red) are shown.

## 2. Correcting for Finite Size Error

As discussed in the main text, despite the large simulation cell we employed (216 atoms - 864 electrons), there remains small, but non-negligible, finite size errors. At the same time, it is not currently possible to perform the ensemble trajectories using a larger simulation cell because of its very large computational cost. We estimated the finite size error by calculating the electronic stopping power for the representative centroid path (see main text) using a simulation cell that is twice as large in the perpendicular and parallel directions relative to the projectile ion trajectory as shown in Figure S2 and S3. In order to correct for the finite size errors as shown in Figures 1 and 2 (main text), a percentage difference (with respect to the original simulation cell) was obtained by averaging the stopping power curves for the two larger simulation cells. This velocity-dependent percentage deviation is then used to obtain the corrected values as shown as the dashed lines in Figure 1 and Figure 2 (main text).

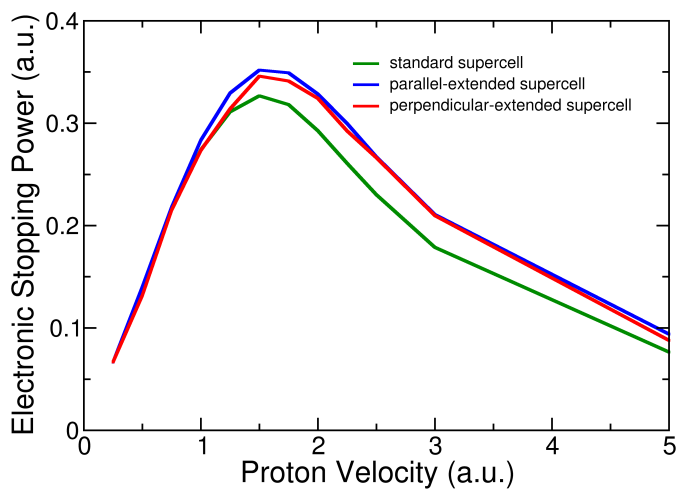

FIG. S2. Electronic stopping power for protons in SiC for three different simulation supercells: The standard cubic supercell containing 216 atoms (green), the supercell doubled (432 atoms) in the direction parallel to the projectile path (blue), and the supercell doubled (432 atoms) in the direction perpendicular to the projectile path (red).

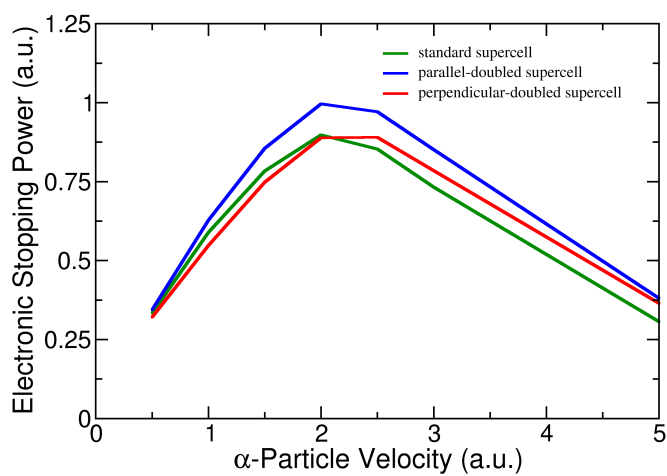

Figure S3. Electronic stopping power for alpha-particles in SiC for three different simulation supercells: The standard cubic supercell containing 216 atoms (green), the supercell doubled (432 atoms) in the direction parallel to the projectile path (blue), and the supercell doubled (432 atoms) in the direction perpendicular to the projectile path (red).

## 3. Comparison between PBE and LDA Exchange-Correlation Potentials

The linear response result by Shukri et. al. [9] is based on LDA exchange-correlation potential. We compared the electronic stopping powers using PBE and LDA exchange-correlation potentials in real-time TDDFT for the centroid path. The LDA yields the stopping power values that are as much as 18 % larger than the PBE results as shown in Figure S4.

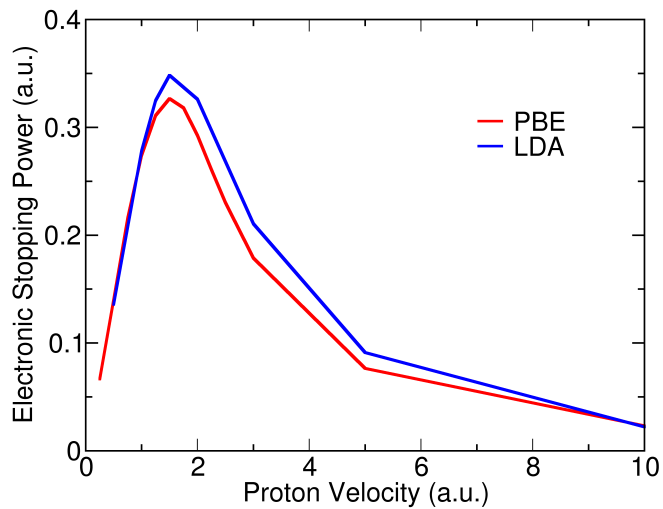

FIG. S4. Electronic stopping powers of protons in SiC using PBE and LDA approximations in real-time TDDFT simulations. No corrections are applied for finite-size errors.

### 4. Charge fluctuations

Using the centroid path, we quantified how the instantaneous charge on the projectile ion changes along the trajectory as a function of ion velocity.

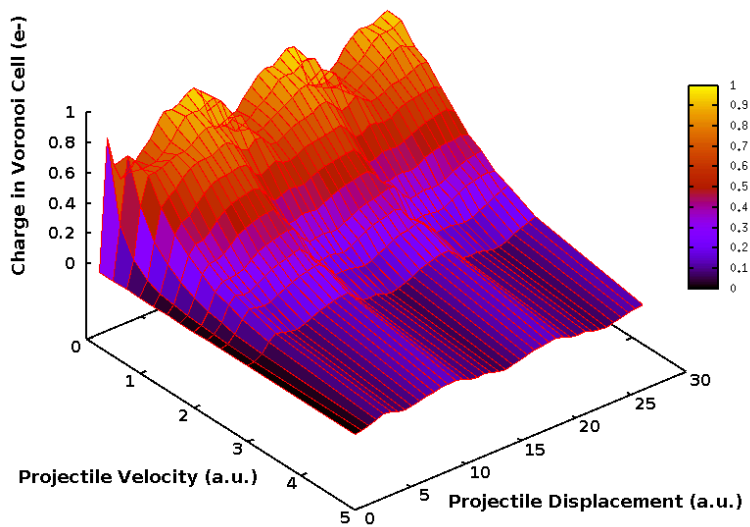

FIG. S5. Instantaneous charge on the projectile proton as a function of ion velocity and the projectile position, according to Voronoi partitioning scheme.

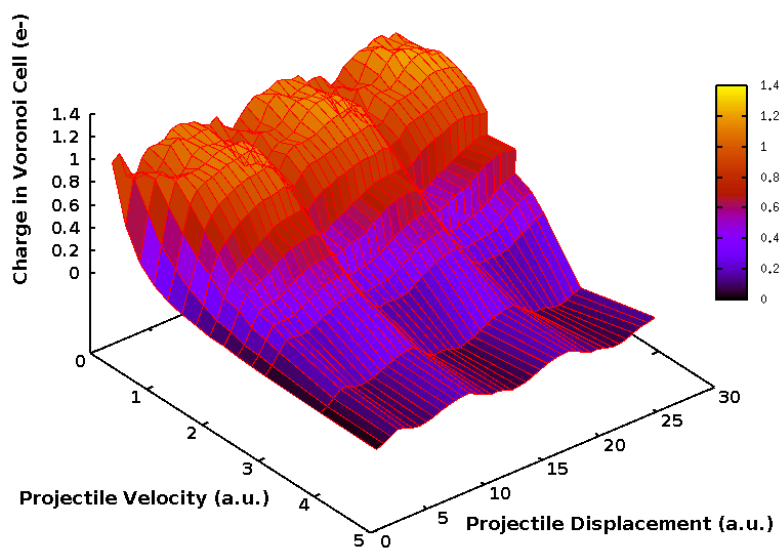

FIG. S6. Instantaneous charge on the projectile α-particle as a function of ion velocity and the projectile position, according to Voronoi partitioning scheme.